\documentclass[fleqn,usenatbib]{mnras}
\usepackage{ulem}
\usepackage{soul}
\usepackage{xcolor}
\usepackage{xspace} 
\newcommand{\noop}[1]{}

\usepackage{newtxtext}

\usepackage[T1]{fontenc}

\RequirePackage{fix-cm}

\usepackage{graphicx}	
\usepackage{amsmath}	
\usepackage{newpxmath}
\usepackage{orcidlink}

\newif\ifrevision
\revisiontrue

\usepackage{orcidlink}

\title[Short title, max. 45 characters]{A variable ADAF disk model for X-ray binary systems}

\author[Chun Xu]{Chun Xu$^{1}$\thanks{E-mail: chun.xuu@shao.ac.cn}\orcidlink{0009-0009-2507-5977}
\\
$^{1}$Shanghai Astronomical Observatory, Chinese Academy of Sciences, Shanghai 200030, China}

\date{Accepted XXX. Received YYY; in original form ZZZ}

\pubyear{\the\year{}}

\begin{document}
\label{firstpage}
\pagerange{\pageref{firstpage}--\pageref{lastpage}}
\maketitle

\begin{abstract}
We propose a variable ADAF disk model for X-ray binary systems. In this model, the accretion flow consists of an outer thin disk and an inner thick turbulent ADAF (renamed TADAF in our later work) torus. The ADAF is turbulent, optically thick, and variable in size. A complete cycle of ADAF contraction, transition to a thin disk, and subsequent re-expansion corresponds to the rapid rise, peak, and decay phases observed in the X-ray outbursts of black hole binaries. This cycle also follows the canonical q-track evolution in the hardness-intensity diagram, going from the low-hard state through the high-soft state and back to the low-hard state. Turbulence in the ADAF, along with second-order Fermi acceleration, naturally generates a power-law particle spectrum that directly corresponds to the power-law X-ray spectra observed in most X-ray binary systems. This model unifies the presence of near-ISCO Fe emission lines with the truncated disk paradigm, as observed in the black hole system GX 339-4. It explains the 35-day period in the neutron star system Her X-1 more effectively through variable ADAF sizes than through a precessing disk. This variable ADAF framework may be extended to explain similar phenomena in active galactic nuclei.
\end{abstract}




\section{Introduction}

X-ray binaries (XRBs) are among the most powerful astrophysical laboratories for studying matter under extreme conditions of gravity, density, and temperature. These systems consist of a compact object, either a black hole (BH) or a neutron star (NS), accreting material from a companion star. The gravitational potential energy released by the infalling gas powers luminous X-ray emission, making them some of the brightest X-ray sources in the sky.

Black hole and neutron star X-ray binaries exhibit remarkable similarities in their observational behavior. Both classes display dramatic, episodic increases in luminosity known as outbursts, during which they transition through distinct spectral states, namely the low-hard state (LHS) and the high-soft state (HSS) and trace out analogous patterns in hardness-intensity diagrams (HIDs; \citealt{ingram2019,Inoue2022}). Given that their central objects are fundamentally different, with a neutron star possessing a solid surface and a black hole having an event horizon, this implies that the observed behavioral similarities likely originate from common features in their accretion flows rather than from the nature of the compact object itself.

The accretion disk model around a central object began with the foundational Shakura–Sunyaev $\alpha$-disk theory \citep{Shakura1973}, and later evolved to include advection-dominated accretion flows (ADAFs) to explain radiatively inefficient accretion \citep{Narayan1994}. \citet{Xu2026jet} 
extended the ADAF model into a universal framework applicable to active galactic nuclei (AGNs), young stellar objects (YSOs), and X-ray binaries, by assuming that the binding energy released during accretion is stored in turbulence rather than radiated away, so that an ADAF can form without requiring a two-temperature plasma.
This newly developed turbulent ADAF (or TADAF) shares many similarities with the old one: they are both geometrically thick, and many of their physical parameters scale with radius in the same way. However, there is one important difference: the new ADAF is optically thick,
whereas the old one is optically thin. The new ADAF model yields jet formation through a funnel of the torus near the central object \citep{Xu2026jet}.

The accretion flow generally consists of an outer thin disk and an inner, thick turbulent ADAF torus. The ADAF region is turbulent, dynamic, and unstable, with its size potentially varying over time. This variability in the inner ADAF torus is closely linked to many observed phenomena in X-ray binaries and active galactic nuclei. In this work, we focus on X-ray binaries and propose a unified disk model applicable to systems hosting both neutron stars and black holes.

This paper is organized as follows. Section 2 presents the construction of a variable disk model. In Section 3, the general properties of the model are compared with observational features. Section 4 offers a detailed discussion of two specific sources, one black hole and one neutron star. Finally, Section 5 provides a general discussion.

\section{The variable disk model}
As discussed in \citet{Xu2026jet}, a standard thin accretion disk typically forms at large radii. Within a range of approximately tens to hundreds of times the radius of the central object, highly energetic turbulence triggers the transition to a turbulent Advection-Dominated Accretion Flow, leading to the formation of a thick disk. Consequently, the accretion disk around a neutron star or black hole generally consists of an outer thin disk combined with an inner thick torus. A similar description is provided by \citet{Inoue2022} in his review.

The geometry of the inner ADAF disk and its connection to the outer thin disk is described as follows. As discussed in Xu (\citeyear{Xu2026jet}; also see Fig. 10.5 in \citealt{Frank2002}), fluid blobs possessing turbulent energy $\eta$  (expressed in units of the local Keplerian energy) traverse oval contours characterized by $e=\eta-1$, where $e$ is the total energy (including the gravity energy) of the fluid blobs. Assuming $\eta$  varies with radius $R$ , the ensemble of all such contours at different radii forms an envelope that defines the ADAF disk shape. For simplicity, we adopt a sigmoid profile (Eq (1)) for $\eta(R)$ in order to emphasize that $\eta$ is a continuous function of R, where $\eta \sim 0$  at large radii (corresponding to Keplerian motion in the thin disk) and $\eta > 0$ at small radii (representing the thick ADAF disk), yielding overall disk morphologies similar to those shown in Figure 1:
\begin{equation}
\eta(R) = e+1 = \frac{\eta_{max}}{1 + \exp\left(k(R - R_c)\right)}
\end{equation}
Here, $R_c$ 
denotes the radius where $\eta = 0.5\eta_{max}$, and k is a scaling parameter controlling the steepness of the sigmoid transition.
In Figure 1, the red envelope (traced by the inner thin curves) delineates the shape of the inner ADAF torus and its smooth connection to the outer thin disk, while the inner thin curves themselves represent the oval equipotential contours at various radii $R$  with their corresponding $\eta(R)$  values. Three values of $\eta_{\rm max}$ are chosen to illustrate how $\eta$ affects the disk size (note: if we change the values of k or $R_c$, the shape of ADAF may also change, but $\eta$ is the most important one). It should be emphasized that the precise ADAF disk geometry is highly sensitive to the functional form of $\eta(R)$ . Our adoption of a simple sigmoid function serves merely to illustrate a plausible envelope shape; the actual relationship between $\eta$  and $R$  may be considerably more complex and warrants further investigation. The half‑opening angle $\theta$ of the funnel is primarily determined by $\eta$ at the innermost radius, through $\sin^2(\theta) = -e = 1-\eta$, where $e$ is the specific total energy of the disk fluid and relates to $\eta$ via $\eta = e+1$ (\S10 in \citealt{Frank2002}). 

\begin{figure}
 \includegraphics[width=\columnwidth]{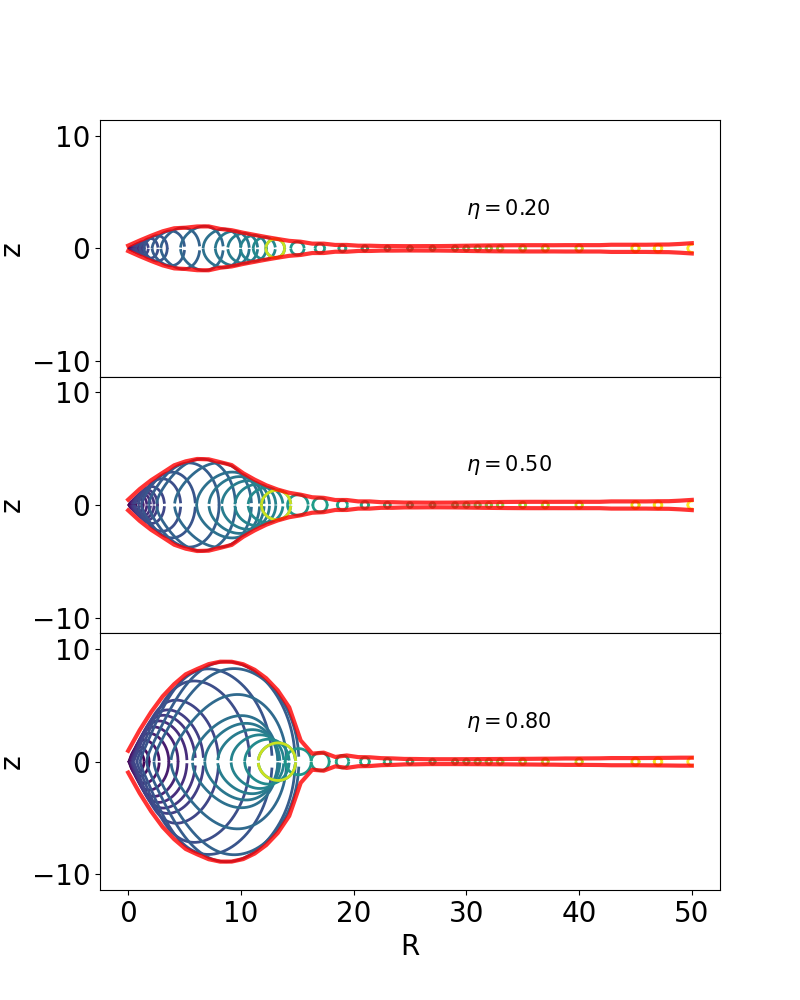}
  \caption{The shape of inner ADAF torus connecting with outer thin disk depicted by red envelope. The bluish thin lines stand for equipotential contour at different radius with different $\eta$ parameter or different level of turbulence. The three panels correspond to three different $\eta_{\max}$ and depict how the ADAF size changes with the maximum $\eta$. Coordinates are in arbitrary scales.}
 \label{fig:ADAF_disk1d}
\end{figure}

Turbulence is inherently dynamic and unstable, so the ADAF is unlikely to remain steady. Its overall shape, size, and internal flows are subject to continuous variation, denoted as $\eta$ for brevity, and as $\eta(R)$ only when the radial dependence is explicitly relevant.
In extreme cases, the innermost region of the disk may oscillate between an ADAF and a thin disk state (when $\eta=0$). In the following chapters, the discussion of the variable ADAF also implies the variability of $\eta$.

The transition of the inner region of the disk between an ADAF and a thin disk requires detailed calculations or simulations. Here, we will discuss this scenario only in a phenomenological manner. Assume that the inner part of the disk is initially in the ADAF state. Since the accretion disk is not static and material continuously accretes onto the central object, the ADAF region also moves inward. If no new turbulence develops in the area previously occupied by the ADAF, then the entire disk—including both the ADAF and the thin disk—will move inward and eventually accrete onto the central object. After a certain period, the entire disk becomes a thin disk, and the ADAF disappears.

However, the thin disk is highly unstable when the local energy release is significant, leading to the reformation of turbulence. As a result, the ADAF is expected to reform, beginning from the innermost region where the local energy release is most intense, and gradually extending outward. Because the disk is continuously accreting inward, the reformation of the ADAF is expected to be slower than its disappearance. Ultimately, a new ADAF forms, and the disk returns to its original configuration with an inner ADAF and an outer thin disk, awaiting the next cycle of transformation.

In reality, the process is more complicated, the ADAF may never fully disappear but instead only oscillate between smaller and larger sizes. Nonetheless, the overall picture remains the same.

Since the ADAF is turbulent, high-speed gas blobs within the turbulence may be ejected and collide above the ADAF surface, forming a corona‑like high‑temperature structure. We refer to this as an ADAF‑corona. Thus, the overall configuration of the accretion flow can be described as follows: an outer thin disk, an inner variable thick ADAF disk, and a dynamic ADAF‑corona situated above the ADAF flow. If the central object is a neutron star, a stable or variable corona may also form above its surface.
If the central object is a black hole, a corona of variable size may form above it, with its extent constrained by both the funnel geometry and the size of the ADAF. The emission from the central corona is expected to be stronger than that from the ADAF‑corona, but their spectra should be similar. In some cases, we do not distinguish between them and simply refer to both as a corona.

The X-ray emissions from X-ray binaries originate from several components: multicolor blackbody emission from the outer thin disk, hard spectral emission from the central corona and ADAF-corona, and emission from the ADAF itself. If the central object is a neutron star, additional contributions include thermal emission from the neutron star surface and hard emission from its corona.

The spectra of the central corona, ADAF-corona or the ADAF itself can be interpreted in the following manner. The ADAF is fully turbulent, and therefore is expected to follow a Kolmogorov energy spectrum, $E(k) \propto k^{-5/3}$, where k denotes the wavenumber of the turbulent eddies. In such a system, the number of particles N possessing a specific energy E is shaped by second-order Fermi acceleration and is governed by a diffusion equation in energy space. This leads to a particle spectrum of the form $N(E) \propto E^{-p}$ , where the power-law index p is approximately 2.0, typically ranging between 1.5 and 3.0 \citep{schlickeiser2002}. This mechanism underlies the power-law X-ray tails observed in most binary systems.

The overall observed spectra of an X-ray binary system require comprehensive simulations that rely on numerous assumptions about the system, including the central object type, accretion rate, instantaneous ADAF size, inclination angle, and the full solutions to the fluid and radiative transfer equations. Such a detailed treatment is beyond the scope of this paper. In this work, we therefore keep our discussion largely phenomenological.

We find that most observed features of X-ray binaries, their light curves and spectra, can be explained within the framework of a variable inner ADAF model, which incorporates emission contributions from the disk, corona, and central object \citep{Inoue2022}. The binary inclination angle may also significantly influence the observed properties, particularly in neutron star systems. In such systems, the ADAF can obscure the line of sight to the central object, while the thermal emission from the neutron star surface itself remains a dominant component.

Before further exploring this model, two important distinctions should be noted. First, the turbulent ADAF in \citet{Xu2026jet} is optically thick, whereas the ADAF in \citet{Narayan1994} is optically thin at low luminosities. This distinction becomes critical when the central source is obscured by the ADAF, optically thick and thin make very different overall spectral appearance. Second, the inner boundary of the ADAF differs depending on the compact object: around a black hole, it extends down to the innermost stable circular orbit (ISCO), approximately $3R_{\rm s}$ (where $R_{\rm s}$ is the Schwarzschild radius $R_{\rm s}=2GM/c^2$). In contrast, around a neutron star, the inner edge is likely more than ten times the stellar radius, as the strong magnetic field may disrupt the innermost region of the ADAF \citep{Xu2026jet}. These differences substantially affect the relative contributions from various emission components.

\section{Overall comparison between model and observations}

The quiescent or low-hard states correspond to periods when the ADAF is large, whereas the outburst or high-soft state is associated with a thin disk or a small ADAF. In a black hole binary system during quiescence, the ADAF extends over a large region, and the thin disk terminates at a large radius. In this state, X-ray emission is primarily produced by the corona and ADAF, which generates relatively hard radiation (i.e., high-energy emission dominates over low-energy emission) and exhibits a power-law tail. A small portion of the X-rays comes from the multicolor blackbody emission of the thin disk, which is weak and soft due to its large inner radius. The combined spectrum appears hard, but the overall luminosity is low, characterizing the low-hard state---this corresponds to the lower branch or ``foot'' of the ``q'' in the q-shaped hardness--intensity diagram (HID; see also Fig. 2 in \citealt{ingram2019}, \citealt{munozdarias2014}, \citealt{belloni2010}). The root mean square (rms) variability amplitude of X-rays is large in this state, as the hard X-rays originate predominantly from the highly variable ADAF and coronae.

When an outburst begins, the ADAF rapidly shrinks and the thin disk extends inward, leading to a sharp increase in multicolor blackbody emission from the thin disk. However, the ADAF-corona takes some time to dissipate, so relatively strong hard emission persists. As a result, the observed spectrum remains relatively hard while the luminosity becomes strong, marking the transition to the high hard state; this corresponds to the upper right corner of the ``q''. During the early phase of the high hard state, the ADAF-corona gradually fades and the hard X-ray flux decreases accordingly. The spectrum softens as emission becomes dominated by the multicolor blackbody component of the thin disk. The system then enters the high-soft state, represented by the top part of the ``q'' from right to left.

Following the burst peak, the ADAF begins to grow again and the thin disk recedes outward. The ADAF corona also redevelops, causing the spectrum to harden once more. This returns the system to the low-hard state, corresponding to the lower right portion of the ``q''. Eventually, the system settles back into quiescence, completing a full cycle. This alternating behavior between ADAF and thin disk dominance accounts for the light curves and q-shaped HID patterns observed in most black hole binary systems \citep{belloni2010, Inoue2022, ingram2019}. In reality, the evolution may be more complex: the ADAF may temporarily expand even during the declining phase, or vice versa, leading to more complicated light curves and HID trajectories.

For neutron star binary systems, the overall structure resembles that of black hole binaries. However, since neutron stars emit persistent blackbody radiation and their strong magnetic fields can truncate the inner accretion disk at a relatively large radius, the radiative contributions from the ADAF and the disk may be reduced compared to those in black hole systems. Additionally, as the ADAF expands, it may partially obscure the line of sight to the neutron star. As a result, the observed light curves and hardness–intensity diagrams (HID) in neutron star systems can differ slightly from those of black hole binaries. Typically, neutron star systems exhibit Z-shaped or atoll tracks in their HID \citep{hasinger1989,ingram2019}.

Black hole and neutron star X-ray binaries (XRBs) exhibit various types of quasi-periodic oscillations (QPOs), as reviewed by \citet{vanderKlis2000, McClintockRemillard2009} and \citet{ingram2019}. In black hole systems, Type-A and Type-B QPOs are associated with relatively high frequencies, while Type-C QPOs occur at lower frequencies. In neutron star systems, flaring branch oscillations (FBOs) and horizontal branch oscillations (HBOs) are observed at higher frequencies, whereas normal branch oscillations (NBOs) appear at lower frequencies. High-frequency QPOs are typically found in the hard spectral state or the upper left region of the “q” diagram, while low-frequency QPOs are located on the right side of the “$q$”.

According to the variable ADAF disk model, high-frequency QPOs correspond to a smaller ADAF region that is closer to the central object, where the mean flow velocity is higher. In contrast, low-frequency QPOs are associated with a larger ADAF region farther from the central object, resulting in a lower mean flow velocity \citep{ingram2019} . 
Figure~22 in \citet{ingram2019} shows that the QPO frequencies increase during the outburst of GX~339-4, which corresponds precisely to the shrinking of the ADAF throughout the outburst phase.
Although the exact mechanism behind QPOs remains an open question, the correlation between their frequencies and the sizes of the ADAF region is well established.

This section provides only a general discussion of the overall behavior of black hole or neutron star X-ray binary systems within the framework of the variable ADAF-thin disk model. A detailed analysis and simulations are required to compare this model with individual objects.

\section{Application to individual sources}

\subsection{Black hole system GX 339-4}

GX 339-4 is a benchmark black hole X-ray binary and one of the most thoroughly studied systems of its kind, offering key insights into black hole accretion states and X-ray spectral evolution. First detected in 1973 \citep{Markert1973}, this transient binary comprises a stellar-mass black hole (with an estimated mass of $\sim 5-10\,\mathrm{M}_{\odot}$) and a low-mass companion star, orbiting each other every $\sim 1.76$ days \citep{Hynes2003}. The source undergoes recurrent outbursts every few years, transitioning through well-defined spectral states: from the low-hard state, where emission is dominated by thermal Comptonization with a photon index of $\Gamma \sim 1.5-1.8$, to the high-soft state, characterized by a prominent thermal disk component and a softer spectrum with $\Gamma \sim 2.5-2.8$ \citep{Miyamoto1991, McClintockRemillard2009}. Notably, GX 339-4 was the first system in which low-frequency quasi-periodic oscillations (QPOs) were securely identified \citep{Motch1983}, and it continues to serve as a key reference for studies of relativistic iron lines and disk reflection \citep{Miller2006, Parker2016}. More recently, observations with \textit{NuSTAR} and \textit{NICER} have refined our understanding of its accretion geometry, revealing intricate connections among disk winds, Comptonization processes, and possible disk truncation in the hard state \citep{zdziarski2020, WangJi2018}.

Many of the observed features of the black hole X-ray binary GX~339-4---including its light curve and the q-shaped track in the hardness--intensity diagram (HID)---are discussed in Section~3. Here, we focus on one of the most debated questions in the field: does the accretion disk truncate at a certain radius, or does it extend all the way to the ISCO? While some studies argue for a truncated disk in the low-hard states \citep{Tomsick2009, Chainakun2021,zdziarski2020}, others, using broad iron line profiles, suggest that the inner disk remains close to the ISCO even at low luminosities \citep{Miller2006, Reis2010}.

The variable ADAF disk model naturally resolves this apparent contradiction. In this model, the accretion disk does extend to the ISCO, but the inner part of the disk is a turbulent ADAF torus whose size changes over time. Support for this comes both from broad iron line features \citep{Miller2006} and from the jet formation requirements proposed by \citet{Xu2026jet}, that jet can form only if the black hole is shrouded by the torus. Evidence for a truncated disk, on the other hand, is typically derived from power spectral density (PSD) or spectral fitting methods \citep{Chainakun2021}, which primarily trace the thin disk component; the inner ADAF region remains largely invisible to these techniques.

\citet{Chainakun2021} reported that the truncation radius increases from $\sim 10\,R_{\rm g}$ to $\sim 55\,R_{\rm g}$ (note: $R_{\rm g}=R_{\rm s}/2$) as the source flux declines toward the end of an outburst, a trend that aligns precisely with the assumptions of the variable ADAF disk framework.

\subsection{Neutron star system Her X-1}

Hercules X-1 (Her X-1) is a neutron star X-ray binary system with a stellar companion, Her HZ. The neutron star exhibits a pulse period of 1.24\,s, and the binary system has an orbital period of 1.7\,days. It also displays an approximately 35-day super-orbital period during which the X-ray flux turns on and off twice, corresponding to the main high state (MHS), short high state (SHS), and low state (LS). The X-ray light curve during the high state shows a rapid rise and slow decay \citep{Tananbaum1972, Giacconi1973, leahy2025, vrtilek2001}. Significant efforts have been made to explain the 35-day period, with the most compelling model involving a precessing warped accretion disk driven by radiation from the central source \citep{Katz1973, Petterson1977, Pringle1996, Wijers1999, leahy2025}. This precessing warped disk model can also account for the pre-eclipse dip phenomenon. However, two important issues remain unresolved. One is the occurrence of the so-called anomalous low state (ALS), during which the normally expected bright main high state in the 35-day cycle almost completely disappears for several cycles \citep{Parmar1985, vrtilek2001}. It is difficult to envisage that simply changing the tilt angle of the warped disk can fully explain this phenomenon. The other issue concerns the pulse profile: the pulse shape differs between the main high state and the short high state, and the pulse peak shifts by about half a pulse period \citep{deeter1998}, which is difficult to fully explain within the framework of the precessing disk model. There is also a free neutron star precessing model that can explain certain specific features of the Her X-1 spectra and periods \citep{Staubert2013,Kolesnikov2022}.

The variable ADAF-thin disk model can explain all the above phenomena, including the pulse profile, 35-day super-period, the pre-eclipse dips by random turbulent edges of the ADAF disk and more. The inclination angle $i$ of Her X-1 is about $85^\circ$ \citep{leahy2025}, so the system is observed nearly edge-on. The neutron star itself exhibits nearly constant X-ray emission regardless of whether the system is in the MHS, SHS, or LS state. The low state corresponds to an extended ADAF, where the torus blocks the line of sight toward the neutron star, resulting in very weak X-ray emission. The turn-on of X-rays occurs when the ADAF transitions into a small ADAF or a thin disk, making the neutron star visible. The decrease in X-ray flux corresponds to the regrowth of the ADAF. As expected from the model, the rise time is usually shorter than the decay time. The difference between MHS and SHS lies only in the size of the ADAF as illustrated in Figure 2. The occurrence of anomalous low state means that throughout that long period, the inner disk remains in large ADAF. After ALS, the 35-day period slightly changes from the one before ALS \citep{still2004,leahy2011}, which is understandable within this variable ADAF model. In fact, even the 35-day period itself is not stable because the cycle is related to instability of turbulence.

\begin{figure}
 \includegraphics[width=\columnwidth]{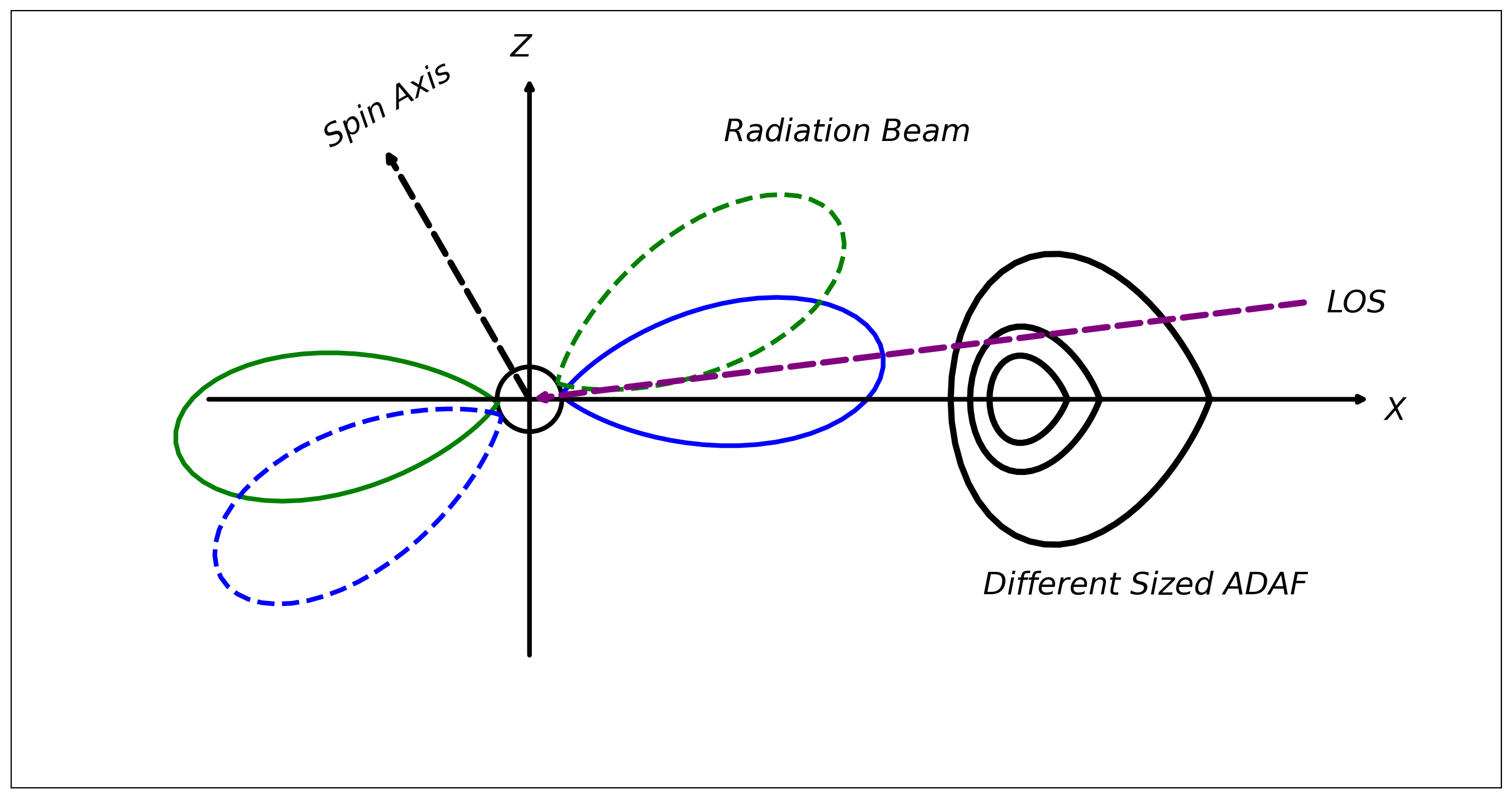}
\caption{Schematic illustration of the Her X-1 system geometry. The accretion disk lies in the XY plane. The spin axis of the neutron star is inclined
from the Z axis, while the two radiation beams are not perpendicular to the spin axis. As a result, emission from both beams can be observed, with alternating high and low flux levels. The primary beam (blue) passes directly through the line of sight (LOS), whereas the secondary beam (green) intersects the LOS only at its edge. The dashed beams indicate the positions of the solid beams after half a rotation period. Variations in the size of the ADAF drive the transitions among the MHS, SHS, and LS states of Her X-1, with the smallest ADAF corresponding to the MHS state.}
\label{fig:neutron_geometry}
\end{figure}

Regarding the pulse phase, the neutron star emits X-rays via two nearly opposite pencil-like beams, with the neutron star's rotating axis at an angle of around $70^\circ$ or so ( note: only a guess rather than from a simulation)  with respect to the opposite beam direction. Both beams pass line of sight (LOS), with the strong beam pointing at the observer while the weaker beam pass above with its edge touch the LOS. Beam's opening angle is about $80^\circ$ as indicated by the width of the pulse profiles \citep{deeter1998}. During MHS, ADAF is small, both beams are visible, so the pulse profile in MHS actually contains two peaks, although the weaker peak is blended with the background. During SHS, the ADAF is large enough to occult the strong beam, leaving only the weak upper beam visible. As a result, the pulse profile in SHS exhibits a weak peak shifted by half a cycle. Fig. 4 in \citet{deeter1998} exactly illustrates this scenario. Moreover, the peak flux of the SHS pulse is indeed lower than the flux at the same pulse phase in MHS, indicating that the weak pulse peak is mixed with the background in MHS. The occultation of the strong beam has also been invoked in warped disk models to explain the pulse shape, but such models require a very complicated combination of the outer and inner parts of the warped disk, carefully tuned to reproduce the observations \citep{blum2000,scott2000}.

According to the proposed scenario, during the SHS state, the strong pulse is expected to be visible when the ADAF does not fully cover the strong beam, whereas the weak pulse may be observable at the beginning or end of the MHS phase, when the strong beam is almost entirely blocked. Such behavior has indeed been observed. Fig. 6 in \citet{scott2000} shows that the strong beam appears during SHS and disappears progressively as SHS evolves and the ADAF size increases (see also several figures in \citealt{deeter1998}). Near the end of MHS, the weak pulse is barely detectable, as illustrated in Fig. 5 of \citet{deeter1998}, around a 35-day phase of approximately 0.23. The peak of the weak pulse occurs near pulse phase 0.7, rather than the standard 0.5. However, the remnant of the strong pulse is also shifted to pulse phase 0.2. This can be understood if the spin axis is tilted relative to the line of sight (LOS), with the neutron star appearing to roll obliquely upward as seen from the LOS, while the ADAF rises from below. The detection of the weak pulse during the late MHS phase remains inconclusive though from the data and requires further observations and more detailed analysis.

Although the exact parameters that determines duration of the 35-day period remain unknown, we suggest that they are related to the size of the ADAF or $\eta$, the average radial velocity of accretion flow, and the dynamic nature of turbulence. As discussed above, the turbulent ADAF-thin disk turning-around model can explain most of the observed features associated with the 35-day period. The mass accretion rate onto the neutron star during the ADAF state is lower than that during the thin disk state, as the ADAF is more extended. This is supported by the following observations that the neutron star pulse period becomes longer than spin-up expected under normal circumstances, e.g., the 1983 ALS lasted for eight 35-day cycles and pulse period extended $20~\mu{\rm s}$, the 1993, 1999 ALSs lasted for four and over ten 35-day cycles and pulse periods extended 12 and $25~\mu{\rm s}$ respectively \citep{coburn2000,vrtilek2001}. This implies a reduced mass accretion rate during the ALS. It is understandable because the ADAF is large during the ALS, causing the inner torus to be more extended and more mass to be lost through outflows, thereby reducing the accretion rate onto the neutron star. According to this reasoning, the model predicts that the neutron star's pulse period varies with the 35-day phase. For example, the neutron star spins up more rapidly during the MHS than during the SHS. This prediction is exactly borne out by observations. \citet{Klochkov2008} found that the spin-up of the neutron star in Her X-1 is correlated with the 35-day period, with the fastest spin-up occurring around the peak of the MHS (see Fig. 4 in \citealt{Klochkov2008}), which corresponds to the smallest ADAF and fastest accretion rate.

Some further comments on the ALS. The 35-day period still exists during ALS, but because the ADAF is large while its variability amplitude is smaller, the neutron star remains blocked by the ADAF even as it varies - just with reduced amplitude. We therefore expect from this reduced amplitude that the 35-day period should be shorter during and around ALS periods. Such a phenomenon has in fact already been observed \citep{still2004,Revnivtsev2026}.

Many discussions on Her X-1 can extend to similar sources such as LMC X-4 which has a 30.5 super-orbital period \citep{lang1981}.

\section{Summary and Discussions}

The advection-dominated accretion flow proposed by \citet{Narayan1994, Narayan1995} has led to a revolutionary understanding of accretion disks since the establishment of the thin disk model \citep{Shakura1973}. \citet{Xu2026jet} introduced turbulence into the ADAF model, extending its applicability to a variety of accretion sources, ranging from active galactic nuclei, X-ray binaries to young stellar objects. The overall structure of the accretion flow consists of an outer thin disk and an inner turbulent ADAF thick disk surrounding the central object. Due to its turbulent nature, and possibly also due to the Papaloizou–Pringle instability (PPI) \citep{papaloizou1984}, the ADAF is dynamic, and its size may vary over time. In fact, the PPI is more likely the mechanism that triggers turbulence, converting a thin disk into a turbulent ADAF or aiding the growth of an existing ADAF. This variable ADAF disk model is directly linked to various observational phenomena in X-ray binary systems.

During a black hole X-ray burst, the initially extended ADAF region of the disk contracts to a very small size, while the thin disk extends inward nearly to the ISCO. Subsequently, the ADAF gradually expands again back to its original size. This cycle is consistent with the light curves, the q-shaped track in the HID diagram, and the spectral evolution observed in most black hole X-ray bursts. If the formation of coronae and outflows by the ADAF torus involves no significant time-delay, the HID would then trace a single inverted-L loop, instead of the characteristic q-shaped track. The time-delay effect is the cause of the q-shape in the HID. Neutron star systems undergo similar cycles, but due to their magnetic field halting the ADAF at a larger radius and the additional contribution from direct neutron star emission, their cycles differ slightly from those of black hole systems.

With this variable ADAF model, we assume that the disk in the black hole system GX~339-4 always extends down to the ISCO region, and only the inner part transitions between an ADAF and a thin disk. This resolves the apparent contradiction between the iron line measurements suggesting the disk reaches the ISCO and the spectral or QPO evidence indicating a truncated disk. 

Using this model, we further find that the 35‑day super-orbital period in the neutron star system Her X‑1 is better accounted for by variations in the size of the ADAF torus (or equivalently, the parameter $\eta$) than by a warped precessing disk, or the free neutron star precession model. Several observational features arise naturally within the variable ADAF framework: the pulse profile difference between MHS and SHS, the presence of the ALS and its associated lower accretion rate, the modulation of the pulse period with the 35‑day phase, and the detection of the complementary pulse shape in the MHS or SHS. By contrast, these same features are difficult to reconcile with either the warped precessing disk model or the free neutron star precession model.

In a subsequent paper, \citet{Xu2026tadaf} applies this variable TADAF framework to re-interpret many of the observed phenomena in the fast-changing microquasar GRS 1915+105, and finds these phenomena to be consistent with the framework.

This model rests on two key assumptions: energy‑bearing turbulence in the accretion disk and variability in the size of the ADAF, both are related to the variable parameter $\eta$. The turbulence is invoked to produce a thick disk necessary for jet formation \citep{Xu2026jet}, while the variation of the ADAF size is originally introduced to account for the X‑ray burst cycle. However, we find that the turbulence itself also contributes to explaining the formation of power‑law X‑ray spectra —-- a ubiquitous feature in XRBs and AGN whose origin remains not fully understood. The varying ADAF size further helps to account for the 35‑day super-orbital period in Her X‑1. In addition, a continuous disk that includes both thin and thick ADAF components extending down to the ISCO naturally resolves the contradiction associated with the truncated‑disk scenario. The observed correlation between QPO frequencies and the size of the ADAF torus also emerges naturally. All these results follow from the model without any additional assumptions. Its ability to account for a broad range of phenomena in X‑ray binary systems thus extends far beyond its original purpose.

This model remains in its early stages and is largely grounded in qualitative assessments. A comprehensive numerical simulation, incorporating key system parameters such as the central object type, accretion rate, ADAF size, and inclination angle, and solving the full set of fluid, magnetic, and radiative transfer equations, will be essential to resolve these issues in future work. On the other hand, the geometry of the variable ADAF has been well established to explain many features of the Her X-1 spectral variability.

The evolution of the ADAF region of the disk is essential for understanding X-ray binary systems. There is a wealth of observational data that can be reanalyzed or reinterpreted within the ADAF framework to study the development of turbulence in ADAF. This variable ADAF framework may also be extended to explain similar phenomena in active galactic nuclei. In a subsequent paper we discuss the origin of Changing-Look AGN which is related to the variable ADAF disk \citep{Xu2026agn}.

\section*{Acknowledgements}
This work is supported by the China Manned Space Program Grant No. CMS-CSST-2025-A19 and Grants allocated for the development of the Multi-Channel Imager and the Integral Field Spectrograph, led by Drs. Zhenya Zheng and Lei Hao respectively, for the Chinese Space-station Survey Telescope (CSST).

\section*{Data Availability}

No new data were created or analysed in this study.




\bibliographystyle{mnras}
\bibliography{example} 





\bsp	
\label{lastpage}
\end{document}